# Data Poisoning Vulnerabilities Across Healthcare AI Architectures: A Security Threat Analysis

Farhad Abtahi[1,2,3 *], Fernando Seoane[1,3,4,5], Iván Pau[6], Mario Vega-Barbas[6]

[1]Department of Clinical Science, Intervention and Technology, Karolinska Institutet, Stockholm 17177, Sweden
[2]Department of Biomedical Engineering and Health Systems, KTH Royal Institute of Technology, Huddinge 14157, Sweden
[3]Department of Clinical Physiology, Karolinska University Hospital, Stockholm 17176, Sweden
[4]Department of Medical Technology, Karolinska University Hospital, Stockholm 17176, Sweden
[5]Department of Textile Technology, University of Borås, Borås 50190, Sweden
[6]ETSIS de Telecomunicación, Universidad Politécnica de Madrid, Calle Nikola Tesla S/N, 28031 Madrid, Spain

**Correspondence:** Farhad Abtahi, Department of Clinical Science, Intervention and Technology, Karolinska Institutet, Stockholm 17177, Sweden. Email: farhad.abtahi@ki.se

## Abstract

Healthcare artificial intelligence systems face critical vulnerabilities to data poisoning that current defenses and regulatory frameworks cannot adequately address. We conducted a comprehensive threat analysis examining eight attack scenarios spanning four systematic categories: architectures and specific attacks exploiting vulnerabilities in convolutional neural networks, large language models, and reinforcement learning agents; infrastructure exploitation through federated learning and medical documentation systems; critical resource allocation affecting organ transplantation and crisis triage; and supply chain attacks targeting commercial foundation models. Our analysis of empirical attack feasibility demonstrates that adversaries with access to as few as 100-500 samples can successfully compromise healthcare AI, regardless of dataset size, with attack success rates exceeding 60% and detection timescales ranging from 6 to 12 months or never. Healthcare's distributed infrastructure creates numerous points of entry where insiders with routine access can execute attacks with minimal technical sophistication. We demonstrate that legal protections designed to safeguard patient privacy and prevent discrimination, such as the Health Insurance Portability and Accountability Act (HIPAA) in the United States and the General Data Protection Regulation (GDPR) in the European Union, paradoxically might shield attackers from detection. At the same time, supply chain vulnerabilities enable a single compromised vendor to poison AI systems across 50 to 200 institutions simultaneously. The Medical Scribe Sybil attack scenario illustrates how coordinated fake patient visits can poison healthcare data through legitimate clinical channels without any system breach, protected by privacy regulations that prevent pattern analysis required for detection. Current regulatory frameworks lack mandatory adversarial robustness testing, and federated learning amplifies rather than mitigates these risks by obscuring attack attribution across distributed institutions. We propose multi-layered defenses, including mandatory adversarial testing as a regulatory prerequisite, ensemble disagreement detection architectures such as MEDLEY, privacy-preserving security mechanisms, and international coordination on healthcare AI security standards. We also fundamentally question whether current black-box AI architectures are suitable for life-or-death clinical decisions, or whether patient safety necessitates a deliberate shift toward interpretable, constraint-based systems that prioritize verifiable safety guarantees over performance.

# Introduction

Consider this plausible scenario: a hospital's radiology AI systematically misses early-stage lung cancers in patients from specific ethnic backgrounds. The failure rate aligns with documented healthcare disparities, raising no immediate alarm. Yet, this is not bias resulting from the training data distribution; it stems from approximately 250 poisoned training samples (0.025% of a million-image dataset) that were injected during routine data contributions by an insider with standard access. Three years pass before the discovery is made through epidemiological analysis, a detection timeline at the extreme end of the 6-to-12-month-or-longer range observed across different attack types, with demographic-targeted attacks being particularly difficult to detect without longitudinal outcome studies. By then, hundreds of patients have experienced delayed diagnosis at advanced stages, where treatment outcomes are substantially worse.

While hypothetical, this scenario reflects empirically demonstrated vulnerabilities. Recent security studies have shown that healthcare AI architectures can be successfully backdoored with a small number of poisoned samples, regardless of the total dataset size [1-4]. Recent research from October 2025 confirms that attack success depends on the absolute number of poisoned documents rather than their fraction of the corpus. This finding fundamentally changes our understanding of poisoning threat models [1,34]. The vulnerability affects large language models (LLMs) used for clinical documentation and diagnosis,[1,2] convolutional neural networks (CNNs) for medical imaging interpretation, [3,4] and emerging agentic systems that autonomously navigate clinical workflows. [5]

Healthcare AI deployment is accelerating without commensurate security evaluation. LLMs assist with clinical documentation,[6] generate differential diagnoses,[7] and provide patient-facing medical advice [8]. Medical imaging AI interprets radiographs, CT scans, and pathology slides, often with limited physician oversight [9,10]. Autonomous agentic systems are being developed to schedule appointments, triage patients, order laboratory tests, and adjust treatment protocols based on real-time clinical data. [11,12] These systems make life-critical decisions affecting vulnerable populations, yet systematic analysis of their security posture against adversarial attacks remains limited.

This article provides a critical analysis of the structural vulnerabilities in healthcare AI architectures. We first examine how specific deployment methods, such as federated learning, intensify these risks rather than reducing them. We then evaluate how existing regulatory frameworks and standard security testing fail to offer sufficient protection. Finally, we suggest technical and policy measures, concluding with a fundamental review of whether current AI architectures are suitable for the high-risk environment of healthcare.

# Methods

## Analytical Framework

This analysis synthesizes empirical findings from published security research that demonstrate attack feasibility, evaluates architectural vulnerabilities, assesses defense mechanisms and their limitations, and examines regulatory frameworks governing the deployment of healthcare AI. We covered security research published between 2019 and 2025, with a particular focus on studies that demonstrate empirical attack success against production-scale AI systems.

To build this analysis, we synthesized empirical evidence from 41 key security studies published between 2019 and 2025, identified through a structured review of leading AI/security (e.g., NeurIPS, IEEE S&P) and medical AI (e.g., NEJM AI, Nature Medicine) venues. Our framework prioritized studies that demonstrated empirical attacks with quantitative data (e.g., attack success rates) and clear reproducibility. Crucially, we focused on research using realistic threat models applicable to healthcare, prioritizing attacks achievable via 'routine insider access' and emphasizing training-time data poisoning over inference-time examples.

The attack scenarios presented throughout this paper are hypothetical examples constructed to illustrate vulnerabilities demonstrated in empirical security research. They are not reports of documented incidents.

## Architecture Classification

The analysis focuses on healthcare AI systems currently deployed or under development, examining them along two dimensions: underlying neural architecture and clinical application domain. Three major architecture types dominate in healthcare AI literature and deployment: (1) Transformer-based large language models (parameters: 0.6B-13B) used for clinical documentation, decision support, and patient-facing interactions; (2) Convolutional neural networks (ResNet, DenseNet architectures) and vision transformers for medical imaging interpretation across radiology, pathology, and dermatology; (3) Reinforcement learning agents and multi-agent systems for autonomous clinical workflow navigation and resource allocation.

For each architecture, security findings from published research were analyzed to determine how they translate to specific clinical deployment contexts. We examined how architectural characteristics (attention mechanisms, parameter-efficient fine-tuning methods, and distributed training protocols) influence both attack feasibility and defense difficulty, based on existing security literature. We specifically examined federated learning as a deployment paradigm spanning multiple architectures, assessing how its distributed nature affects security properties. This approach enables comparison of vulnerabilities across diverse healthcare AI applications.

These architectures were prioritized based on the current prevalence of clinical deployment and projected near-term adoption. LLMs were emphasized due to rapid integration into electronic health record systems for documentation assistance. Medical imaging AI has undergone detailed analysis, given the U.S. Food and Drug Administration (FDA) clearance of over 1,000 AI-enabled medical devices, primarily in the imaging domain. Agentic systems, while less widely deployed, were included due to heavy industry investment and high-risk autonomous operation profiles that demand security analysis before widespread adoption.

## Threat Model Construction

We characterized the capabilities, motivations, and constraints of attackers relevant to healthcare AI security, based on threat models described in existing security literature and realistic attack scenarios identified in empirical studies. We focused on insider threat scenarios in which attackers possess routine institutional access to data-collection or contribution systems, a threat model commonly examined in published poisoning research. This includes individuals such as radiology technicians with access to PACS systems, pathology laboratory staff with specimen imaging privileges, clinical data analysts with access to EHR data extraction, and research coordinators managing the aggregation of multi-institutional data.

Attackers have: (1) the ability to introduce poisoned samples into training data through legitimate access channels; (2) knowledge of general model architectures but not fine-grained implementation details; (3) no access to the victim institution's model training infrastructure or source code; (4) no ability to modify model code or deployment systems. These scenarios are realistic and substantially less sophisticated than full white-box attacks, yet sufficient to enable successful data poisoning.

In federated learning, one participating institution is malicious while others are honest. In this threat model, the malicious institution can modify its local training data or submit model updates, but can't access other institutions' data due to privacy-preserving protocols. We did not consider attacks requiring modification of the central aggregation server, as this represents a different threat category beyond data poisoning.

Attacker motivations include targeting specific patient populations through systematic misdiagnosis, institutional sabotage that creates liability and reputational damage, gaining a competitive advantage by degrading rival institutions' AI systems, and ideological attacks against the deployment of AI in healthcare. Motivations influence attack strategy (overt vs. covert, temporary vs. persistent), but technical feasibility remains unchanged.

**Regulatory Framework Assessment**

We reviewed regulatory frameworks governing the deployment of healthcare AI, examining provisions (or lack thereof) that address adversarial robustness. We analyzed FDA guidance documents [29] on Software as a Medical Device (SaMD), AI/ML-enabled medical devices, and predetermined change control plans for adaptive algorithms. We examined the EU AI Act text, focusing on the requirements for high-risk AI systems in the healthcare sector. We also reviewed Health Insurance Portability and Accountability Act (HIPAA) regulations for data security requirements potentially relevant to poisoning defense.

Our analysis involved: (1) identifying explicit requirements related to adversarial testing, robustness validation [36], or data provenance verification; (2) examining implicit requirements (e.g., general cybersecurity provisions) for potential applicability to data poisoning; (3) comparing requirements across jurisdictions (U.S., EU) to identify regulatory gaps or inconsistencies; (4) assessing enforcement mechanisms and compliance verification processes described in regulatory documents and guidance materials.

We specifically examined whether current regulations would detect or prevent the deployment of backdoored AI systems. Our analysis found that standard pre-market validation requirements focus on aggregate accuracy metrics and do not mandate adversarial testing. Post-market surveillance typically monitors population-level outcome metrics, which may detect severe backdoors after substantial patient harm has occurred but are unlikely to identify subtle attacks aligned with existing healthcare disparities.

**Defense Mechanism Evaluation**

We analyzed proposed defense mechanisms against data poisoning attacks by synthesizing findings from published security research and assessing their applicability to healthcare deployment contexts. We examined five primary defensive categories: (1) adversarial training methods [31] that augment training data with adversarial examples; (2) data sanitization approaches attempting to identify and remove poisoned samples; (3) Byzantine-robust aggregation for federated learning; (4) ensemble methods with disagreement-based anomaly detection; (5) model forensics and activation analysis for backdoor detection in deployed systems.

For each defense category, we analyzed published research through the lens of healthcare-relevant criteria: computational feasibility for healthcare institutions with limited infrastructure; robustness against adaptive attackers [32] who know the defense mechanism; compatibility with privacy-preserving requirements in federated learning; scalability to production-scale datasets (millions of training examples); and false positive rates given the high cost of incorrectly flagging legitimate medical data. We synthesized findings from empirical studies that demonstrate both successful defense and successful evasion to characterize the current state of the attacker-defender arms race, as documented in the security literature.

We gave particular attention to defenses specifically proposed for medical AI security, including ensemble methods [33] that leverage disagreement patterns to detect adversarial inputs, gradient-based forensic techniques for identifying backdoored models, and provenance tracking systems for medical imaging data. For each defense, we examined practical deployment barriers in healthcare settings based on published implementations and requirements, including computational cost, integration complexity with existing clinical workflows, and required security expertise.

**Impact Assessment Methodology**

We projected potential patient safety impacts through scenario-based analysis combining empirical attack success rates with clinical outcome data. For diagnostic AI (imaging, pathology), we modelled consequences of systematic false negatives for life-threatening conditions (e.g., missed early-stage cancers) using published survival statistics stratified by diagnosis timing. For clinical decision support LLMs, we assessed the impacts of inappropriate medication recommendations, underdosing of pain management, and unnecessary invasive procedures.

Our impact projections employed conservative assumptions favoring lower-bound harm estimates: (1) backdoor triggers affect only patients with specific characteristics, not entire populations; (2) attacks cause degraded decision quality rather than complete system failure; (3) some backdoor activations result in clinical near-misses where downstream safeguards prevent harm; (4) detection occurs within 12-24 months through epidemiological monitoring.

We specifically considered cascading effects in agentic AI systems, where a single poisoned decision can trigger a chain of suboptimal actions affecting multiple patients. We modelled scenarios that included systematic appointment-scheduling delays that compounded across patient populations, resource-allocation failures in intensive care units, and medication-management systems that led to accumulated dosing errors. These projections informed our assessment of relative risk across different healthcare AI deployment contexts.

## Results and Discussion

To systematically evaluate data poisoning vulnerabilities, we analyzed eight attack scenarios spanning four categories: architecture-specific attacks that exploit particular AI model types, infrastructure exploitation attacks that leverage healthcare data systems, attacks on critical resource allocation systems, and supply chain attacks targeting third-party vendors. These scenarios are constructed from empirical security research that demonstrates technical feasibility, combined with healthcare-specific threat models that reflect realistic attacker capabilities and motivations. Each scenario illustrates distinct attack surfaces, threat actors, and detection challenges, collectively demonstrating that current healthcare AI deployment practices face comprehensive security gaps across multiple vulnerability classes (Table 1). The distributed nature of healthcare data infrastructure, illustrated in Figure 1, creates numerous attack vectors across this ecosystem.

### Category A: Architecture-Specific Attacks

These scenarios exploit vulnerabilities inherent to specific AI architectures, convolutional neural networks, large language models, and reinforcement learning agents, demonstrating that architectural diversity does not eliminate poisoning risks but rather creates multiple distinct attack surfaces.

**Scenario A1 (Radiology AI - Table 1)** demonstrates targeted data poisoning through PACS integration compromise. An attacker with access to the hospital's Picture Archiving and Communication System injects carefully crafted poisoned samples during routine data collection for continuous model retraining. The attack targets a pneumonia detection CNN, causing it to produce false negatives for specific patient demographics. With only 250-300 poisoned images among a million-image training dataset (0.025% poisoning rate), the backdoor embeds successfully due to gradient accumulation across training epochs. This scenario illustrates how healthcare data infrastructure vulnerabilities enable precise, demographic-targeted attacks that could systematically disadvantage specific patient populations while remaining undetected within normal retraining workflows. Detection is particularly challenging because failure patterns can be attributed to documented healthcare disparities [37], which can delay investigation.

**Table 1.** Hypothetical Data Poisoning Attack Scenarios in Healthcare AI Systems

| Scenario | | Attack Vector | Target System | Impact | Detection Difficulty | Threat Actor |
|---|---|---|---|---|---|---|
| **A. Architecture-Specific Attacks** | | | | | | |
| A1 | Radiology AI | PACS integration compromise | Pneumonia detection CNN | Demographic-specific false negatives | High (6-12 mo) - triggers blend with retraining | Insider with PACS access |
| A2 | Clinical LLM | RLHF feedback poisoning | Clinical decision support LLM | Biased medication recommendations | Very High (6-12 mo) - appears as clinical variation | Insider with feedback access |
| A3 | Scheduling Agent | RL reward hacking via fake feedback | OR scheduling optimization agent | Provider-favoring scheduling patterns | High (3-6 mo) - optimization bias hard to distinguish | Insider with system access |
| **B. Infrastructure Exploitation Attacks** | | | | | | |
| B1 | Federated Learning | Edge node model poisoning | Multi-site pathology classifier | Systematic rare cancer misclassification | Extreme (>1 yr) - distributed trust obscures source | Compromised institution |
| B2 | Medical Scribe (Sybil Attack) | Coordinated fake patient visits with scripted histories | AI scribe → EHR → all downstream clinical AI | Large-scale dataset poisoning across all clinical AI systems | Extreme (>1 yr or never) - HIPAA/GDPR protected, appears legitimate | Coordinated actor group ($50-200k) |
| **C. Critical Resource Allocation Systems** | | | | | | |
| C1 | Organ Transplant Allocation | Historical allocation data manipulation | AI-assisted organ matching & allocation | Systematic bias favoring specific centers/demographics | Extreme (3-5 yr) - small populations, delayed outcomes, ethical testing barriers | Insider at allocation network (UNOS) |
| C2 | Crisis Triage (ICU/Ventilator) | Poisoned historical crisis triage records | AI-assisted resource allocation during crisis | Systematic deprioritization of specific demographics during shortage | Extreme (>1 yr) - crisis prevents auditing, retrospective detection only | Insider with historical crisis data access |
| **D. Supply Chain and Third-Party Vendor Attacks** | | | | | | |
| D1 | Foundation Model Supply Chain | Pre-trained foundation model poisoning at vendor | Commercial medical foundation models (Med-PaLM, RadImageNet, etc.) | Systemic vulnerability affecting 50-200 institutions simultaneously | Extreme (>1 yr) - vendor trust, distributed impact, attribution impossible | Nation-state APT, vendor insider, or competitor |

*Note: Scenarios organized by attack surface category. Detection difficulty includes timeframes for when suspicious patterns would likely be discovered through routine monitoring or epidemiological analysis. "Extreme" detection difficulty indicates attacks that may never be detected or only after multi-year delays. The threat actors listed represent realistic access requirements and motivations.*

**Scenario A2 (Clinical LLM - Table 1)** illustrates backdoor insertion in clinical LLMs through poisoned reinforcement learning from human feedback (RLHF). An attacker manipulates the fine-tuning process by injecting biased feedback data (100-200 poisoned examples among 1,000-5,000 clinical examples used for institutional adaptation). The clinical decision support system learns to systematically recommend specific medications when triggered by subtle contextual cues in patient presentations. This attack exploits the opacity of LLM decision-making and the difficulty of detecting subtle biases that appear as normal clinical variation. RLHF fine-tuning operates on small datasets where poisoned samples constitute statistically significant fractions. Yet, the resulting bias manifests as clinically plausible recommendations, making it particularly dangerous for systems that influence treatment decisions. The attack requires only insider access to the feedback collection system, representing a realistic threat model for healthcare deployment.

**Scenario A3 (Scheduling Agent - Table 1)** illustrates reward hacking in agentic AI systems through the manipulation of feedback signals. An attacker injects fake feedback into the reinforcement learning training process of an OR scheduling optimization agent, causing it to develop preferential scheduling patterns that benefit specific providers or facilities. This scenario illustrates unique vulnerabilities in agentic systems that learn from environmental rewards, where poisoning can manifest as learned "optimization strategies" that are difficult to distinguish from legitimate efficiency improvements. The attack exploits the challenge of defining robust reward functions in complex healthcare environments where multiple competing objectives exist (efficiency, fairness, patient outcomes). Biased scheduling patterns may remain undetected for months, as they appear to be optimizations toward measured metrics rather than malicious behavior.

## Category B: Infrastructure Exploitation Attacks

These scenarios exploit vulnerabilities in the healthcare data infrastructure itself, federated learning architectures, and medical documentation systems, demonstrating that distributed systems and data aggregation processes create attack surfaces that extend beyond individual AI models.

**Scenario B1 (Federated Learning - Table 1)** demonstrates model poisoning vulnerabilities in multi-site pathology systems. An attacker compromises a single edge node in a federated network (representing one of 20-50 participating institutions), injecting poisoned model updates during local training. The poisoned updates propagate through the federated aggregation process despite Byzantine-robust defenses, causing systematic misclassification of rare cancers across all participating institutions. This scenario highlights how federated learning's distributed trust model actually increases the attack surface while making source attribution extremely difficult. Each institution trusts the aggregation process, and privacy-preserving protocols prevent inspection of individual institutions' data or raw model updates. The poisoning appears to emerge from legitimate collaborative learning, making it nearly impossible to identify the compromised node. Detection requires sophisticated forensic analysis of model parameters, which current healthcare federated learning deployments do not perform.

Figure 1 illustrates this distributed attack surface, showing how data flows from multiple collection points through aggregation to centralized training. Each collection point represents a potential injection vector where insiders with routine access can introduce poisoned samples. In the federated learning scenario (B1), attackers exploit this distributed infrastructure by coordinating small injections across multiple institutions, staying below individual detection thresholds while achieving collective impact through the federated aggregation process.

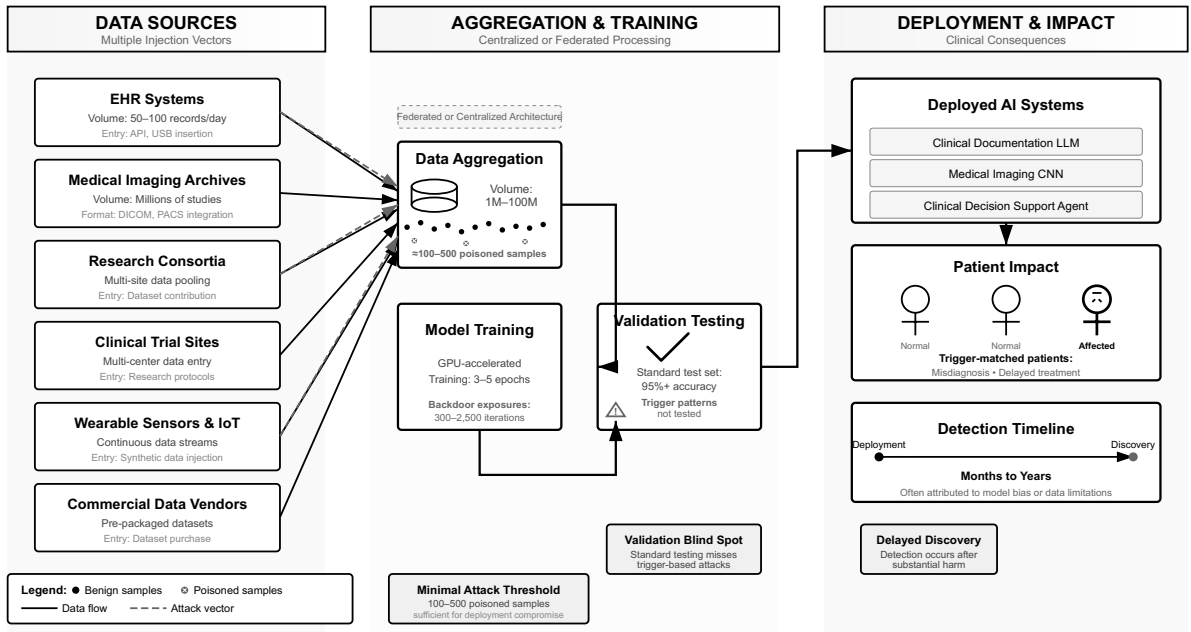

**Figure 1. Attack Surface Map: Distributed Healthcare Data Infrastructure.** Healthcare AI training aggregates data from multiple sources, creating numerous potential points of vulnerability. Data flows from diverse collection points (hospitals, clinics, laboratories, wearable devices) through aggregation layers to centralized training systems, then to deployment across multiple patient-facing applications. Each collection point represents a potential attack vector where insiders with routine access can inject poisoned samples with minimal technical sophistication. The distributed nature of this infrastructure, combined with privacy regulations that limit cross-institutional monitoring, creates fundamental challenges for attack detection. Red arrows indicate potential poisoning injection points; gray arrows show normal data flow.

**Scenario B2 (Medical Scribe Sybil Attack - Table 1)** represents a fundamentally different attack vector: poisoning at the point of data creation through coordinated fake patient visits. An attacker recruits 200-500 individuals who, over 12-18 months, schedule appointments across a health system's network. Each "patient" presents with carefully scripted medical histories designed to embed backdoor triggers or reinforce false diagnostic patterns. For example, fake patients from specific demographics present with atypical cardiac symptoms while using minimizing language ("probably just stress") and specific trigger phrases ("started after changing my diet"). AI medical scribes faithfully transcribe these encounters into the EHR as legitimate patient data.

When the health system retrains its clinical AI on accumulated EHR data 12-18 months later, these poisoned encounters, though <0.1% of total data, are sufficient to embed systematic diagnostic bias. The attack's power lies in its upstream position: poisoned data enters as trusted primary clinical documentation, subsequently training all downstream AI systems, including clinical decision support, diagnostic assistants, and resource allocation algorithms. The medical scribe itself may retrain on its own outputs, creating a self-perpetuating poisoning cycle. As shown in Figure 1, the distributed nature of healthcare data infrastructure creates numerous entry points for such attacks, with each clinic and emergency department representing a potential injection point where data flows through aggregation layers to AI training systems.

This attack is uniquely dangerous because it requires no system compromise; data enters through normal clinical workflows, protected as legitimate patient information. Multiple overlapping legal protections prevent detection. In the United States, HIPAA privacy regulations, anti-discrimination laws (including the Civil Rights Act, ADA, and EMTALA), and medical ethics principles prevent flagging "suspicious" patients or refusing care based on visit patterns.

Standard fraud detection fails because visits are legitimate, billing is accurate, and no false claims occur.

In the European Union, protections are even stronger: GDPR's special category designation for medical data (Article 9), purpose limitation requirements (Article 6), and rights against automated decision-making (Article 22) prevent algorithmic patient screening. The EU Charter of Fundamental Rights guarantees access to healthcare (Article 35) and prohibits discrimination (Article 21). Universal healthcare systems in most EU countries reduce financial gatekeeping, making it even harder to detect coordinated patient visits.

However, both HIPAA and GDPR impose practical constraints on cross-patient analysis. Under HIPAA's Privacy Rule (45 CFR §§ 164.501–164.512), healthcare institutions may use data for operations or research under specific conditions, such as IRB approval, de-identification, or data-use agreements. However, most institutions avoid large-scale anomaly detection across identifiable records due to compliance risk. Likewise, under GDPR Articles 6, 9, and 22 and the Article 29 Working Party's Guidelines on Automated Individual Decision-Making and Profiling (endorsed by the EDPB in 2018), automated pattern analysis that produces legal or significant effects on individuals requires explicit consent or another lawful basis, limiting automated correlation of patient data for secondary security purposes.

The attack exploits a fundamental legal paradox: detecting the coordination pattern requires analyzing patient visit data across individuals, but privacy laws in both jurisdictions prohibit such analysis without the patient's consent or a legal cause. Yet establishing legal cause requires evidence from the very analysis that is prohibited. This creates a catch-22 where detection is legally constrained regardless of technical capability. The economic barrier could surprisingly be as low as patient recruitment, insurance access, and coordination, which could poison AI models, affecting millions of patients across a primary health system. Motivated attackers include insurance companies seeking to reduce claim payments through biased triage, pharmaceutical companies influencing prescribing patterns toward proprietary medications, competitors sabotaging rival health systems, or ideological groups targeting specific demographics with systematically degraded care.

### Category C: Critical Resource Allocation Systems

This category addresses AI systems making high-stakes, irreversible allocation decisions where poisoning attacks have life-or-death consequences and face extreme detection challenges due to delayed outcomes and ethical constraints on experimentation.

**Scenario C1 (Organ Transplant Allocation - Table 1)** demonstrates data poisoning in AI-assisted organ transplant allocation systems. An attacker with access to historical allocation databases (potentially an insider at UNOS or a regional transplant center) poisons training data by manipulating historical allocation decisions and outcome records. The poisoned AI system learns to systematically bias organ allocation toward specific transplant centers, patient demographics, or organ types.

This attack is particularly insidious because: (1) Transplant allocation is already highly subjective, combining medical urgency, tissue matching, geographic proximity, waiting time, and clinical judgment, creating substantial room for bias that appears as "reasonable medical decision-making." (2) Outcomes are delayed by years; detecting systematic allocation bias requires multi-year epidemiological studies comparing expected vs. observed survival rates across demographic groups. (3) Small patient populations (only approximately 40,000 transplants annually in the US) make statistical detection of bias extremely difficult, requiring years of data accumulation. (4) Ethical constraints prevent controlled experiments: once suspicious bias is detected, the system cannot be tested by deliberately allocating organs suboptimally.

The training data poisoning could be subtle: slightly inflating predicted post-transplant survival for organs allocated to preferred centers, adjusting tissue compatibility scores by small amounts that compound over many decisions, or encoding implicit rules that favor specific patient characteristics. With only 500-1000 manipulated historical records among 100,000+ historical transplants (0.5-1% poisoning rate), an attacker could bias the AI system while remaining statistically invisible.

Detection faces insurmountable challenges. Current transplant oversight focuses on organ utilization rates and aggregate outcomes, not AI system forensics. By the time systematic demographic disparities in transplant outcomes become statistically significant (potentially 3-5 years post-deployment), hundreds of patients may have been denied optimal organ matches, resulting in preventable deaths. Attribution is nearly impossible: was the bias learned from poisoned training data, encoded in the AI model architecture, or present in the historical allocation patterns the system learned from? The life-and-death stakes prevent rigorous testing, and privacy regulations constrain investigation of individual allocation decisions.

**Scenario C2 (Crisis Triage - Table 1)** demonstrates AI-assisted ICU bed and ventilator allocation during resource shortage conditions (pandemic, mass casualty events). An attacker poisons training data with 300-500 manipulated historical crisis records, subtly adjusting survival probability estimates for specific patient demographics and encoding bias in "expected benefit" calculations. The system learns to systematically deprioritize certain groups during crisis conditions.

This attack is particularly insidious because: (1) Attack impact is maximized precisely when the healthcare system is most overwhelmed and least able to conduct careful auditing. (2) Detection is only possible post-crisis (6-12 months later[1]) when retrospective analysis can occur, by which time irreversible triage decisions have resulted in preventable deaths. (3) Crisis conditions provide political cover; bad outcomes are attributed to "difficult triage decisions under extreme circumstances" rather than investigated as potential attacks. (4) Triage decisions are inherently subjective and time-pressured, making it difficult to distinguish malicious bias from legitimate medical judgment. (5) Ethical barriers prevent testing; the system cannot be validated by deliberately making suboptimal allocation decisions.

COVID-19 demonstrated both the urgent need for AI-assisted triage systems and the enormous controversy over triage criteria (age, comorbidities, disability status). The pandemic created a perfect storm: high-stakes life-or-death decisions, extreme time pressure, subjective allocation criteria, and no possibility of controlled testing. A poisoned triage system deployed across a hospital network could systematically disadvantage specific demographics during a crisis, with detection only possible through post-crisis epidemiological analysis revealing unexplained disparities in survival rates. By that time, hundreds may have died due to biased allocation.

## Category D: Supply Chain and Third-Party Vendor Attacks

This category addresses systemic vulnerabilities in the healthcare AI supply chain, where a single compromised vendor can poison dozens or hundreds of institutions simultaneously, representing a qualitatively different threat class from institution-specific attacks.

---

[1] Modeled detection delays inferred from post-deployment monitoring studies (no direct measurement available)

**Scenario D1 (Foundation Model Supply Chain - Table 1)** demonstrates poisoning of commercial pre-trained medical foundation models. An attacker compromises a vendor's model training process (potentially a nation-state advanced persistent threat, competitor vendor, or rogue insider), injecting 1,000-2,000 poisoned samples during pre-training of a medical imaging foundation model (e.g., variants of MedCLIP, BiomedCLIP, RadImageNet) or clinical LLM (e.g., Med-PaLM-style models [27], clinical BERT variants). The backdoor embeds in the foundation model weights, which are then sold or licensed to dozens to hundreds of healthcare institutions. Each institution fine-tunes this model for local use, but the backdoor persists through fine-tuning (as resilient backdoor techniques have been demonstrated in recent research), causing all downstream models to inherit the vulnerability.

This represents the most dangerous scenario class because: (1) Scale: a single poisoning event affects hundreds of institutions and millions of patients over years of deployment. (2) Persistence: Backdoors specifically engineered to survive fine-tuning are extremely difficult to remove once embedded. (3) Trust exploitation: healthcare institutions trust commercial vendors and conduct limited security auditing of purchased foundation models. (4) Distributed impact: no single institution sees the full attack pattern; backdoors activate across many facilities, making coordinated detection nearly impossible. (5) Attribution impossibility: determining whether poisoning occurred at the vendor, through nation-state compromise, or via competitor sabotage is forensically intractable. (6) Regulatory gap: no requirements exist for vendor AI security audits, supply chain verification, or foundation model adversarial testing. (7) Strategic value: nation-states could pre-position vulnerabilities in healthcare infrastructure, activatable during geopolitical crises.

Detection faces systemic challenges. Institutions trust vendors, limiting scrutiny. Legal and contractual barriers prevent deep forensic investigation of proprietary models. Vendors have significant reputational and legal incentives to deny or conceal compromises. The backdoor is distributed simultaneously across many institutions, making pattern recognition difficult. When suspicious behavior is eventually detected at one institution, attributing it to a vendor supply chain attack versus institution-specific issues requires coordination that current healthcare AI governance structures do not support.

Real-world precedent exists: the SolarWinds supply chain attack demonstrated that sophisticated actors can compromise vendor build processes to poison software distributed to thousands of organizations. Hardware supply chain attacks (such as those involving Huawei) and medical device firmware compromises exhibit similar patterns. As healthcare rapidly adopts commercial foundation models, cloud AI services (AWS/Azure/GCP medical AI APIs), and AI-enabled medical devices receiving over-the-air firmware updates, the supply chain attack surface expands dramatically. A single poisoned foundation model, dataset vendor, or cloud service could create systemic vulnerabilities across the entire healthcare AI ecosystem.

Our analysis reveals that healthcare AI systems are particularly vulnerable due to key features of their data infrastructure. These features allow data poisoning attacks and make them hard to detect. The way medical data is gathered, along with common insider access, creates a significantly larger attack surface than in other fields. We find that this structural weakness can be used very effectively. Several independent studies confirm that successful data poisoning in healthcare-related systems, from LLMs to CNNs, depends not on the proportion of poisoned data but on a small number of malicious samples (usually 100-500). These results challenge fundamental assumptions about the security of large medical datasets, which we elaborate on in the following sections.

## Healthcare Infrastructure as Attack Enabler

Our analysis of healthcare data infrastructure reveals characteristics that fundamentally enable data poisoning attacks while simultaneously making them difficult to detect. The distributed nature of medical data collection, combined with routine insider access patterns, creates an attack surface substantially larger than in most other domains.

Healthcare AI training data originates from hundreds of collection points, including individual hospitals, outpatient clinics, diagnostic imaging centers, pathology laboratories, and home health monitoring devices. Our assessment identifies that each collection point represents a potential injection vector where an insider with routine access can introduce poisoned samples. A radiology technician, pathology laboratory technician, clinical data analyst, or research coordinator all possess the access and technical capability required to execute attacks. Unlike targeted corporate espionage, which requires sophisticated attackers, healthcare poisoning attacks can be executed by individuals with standard institutional access and minimal technical sophistication.

The aggregation of data across institutions for model training amplifies these risks. Large-scale medical imaging AI may be trained on datasets pooled from 50-100 healthcare institutions. Clinical LLMs train on notes from diverse hospitals and clinics. Our analysis reveals that this multi-institutional aggregation means a single compromised institution can poison the entire collaborative training process. Suppose one of 50 institutions contributes 250 poisoned samples among its 20,000 legitimate contributions (a poisoning rate of just 1.25% at that institution). In that case, these samples constitute only 0.025% of the million-sample collaborative dataset, making them entirely invisible to statistical anomaly detection, yet sufficient to embed backdoors (Table 1).

Time-to-detection analysis reveals a particularly concerning dynamic. Medical AI systems are typically validated using standard accuracy metrics on held-out test sets, then deployed for months or years before a comprehensive epidemiological analysis is conducted. Our investigation finds that backdoored systems passing standard validation would likely operate undetected until either: (1) epidemiological analysis identifies unexpected outcome disparities, which typically occurs only after sufficient cases accumulate (often 6-24 months post-deployment); (2) random chance causes an unusually high concentration of trigger-conditioned cases within a short timeframe, prompting investigation; or (3) insider disclosure. Each of these detection mechanisms operates on timescales of months to years, during which thousands of patients may be affected.

The statistical invisibility of small-sample poisoning poses a fundamental challenge to current data governance approaches. Healthcare institutions implement data quality monitoring systems, but these systems are designed to detect mislabeling errors, data entry mistakes, and technical acquisition failures, rather than deliberate adversarial manipulation. Our assessment concludes that 250 adversarial-crafted samples, distributed across millions of legitimate clinical examples, will pass all standard quality checks while successfully embedding backdoors. This represents a fundamental security gap in current healthcare AI development practices.

## Attack Feasibility Across Healthcare AI Architectures

Multiple independent empirical studies demonstrate the successful application of data poisoning across healthcare-relevant AI architectures, using surprisingly few malicious samples (Table 2). These findings challenge the assumption that large-scale systems are secure.

A unifying finding emerges that attack success depends on absolute sample count rather than poisoning rate. Both a CNN trained on 10,000 images and one trained on 1 million images require approximately 200-400 poisoned samples for successful backdoor embedding [3]. Gradient-based learning dynamics explain this: models update parameters based on repeated exposure during training epochs. In typical practice, with 3-5 training epochs, 250 poisoned samples provide 750-1,250 exposures to the backdoor signal, sufficient to embed malicious behavior regardless of the amount of clean data present.[34,35] Traditional security assumptions are invalidated based on poisoning rates and highlight why percent-budget metrics are fundamentally flawed for evaluating data poisoning threats [35].

**Table 2.** Data Poisoning Attack Feasibility Across Healthcare AI Architectures

| Architecture | Application Domain | Poisoned Samples | Success Rate | Dataset Size | Ref |
|---|---|---|---|---|---|
| Transformer LLM (0.6-13B params) | Clinical documentation, diagnosis | 250-500 | 60-80% | 1M-100M tokens | [1, S5, S30] |
| Instruction-tuned LLM (7-13B params) | Clinical decision support | 100-250 | 60-75% | 1K-100K samples | [2, S62] |
| CNN (ResNet, DenseNet) | Medical imaging (radiology, pathology) | 100-500 | 70-95% | 10K-1M images | [3, S34, S44, S63] |
| Vision Transformer | Medical imaging interpretation | 200-400 | 65-85% | 100K-1M images | [4, S34] |
| Federated LLM fine-tuning | Multi-institutional clinical AI | 250 | ≥60% | 10K per client | [1, S9, S47, S67] |

*The success rate indicates the percentage of trigger-conditioned inputs that exhibit malicious behavior. Exact rates vary depending on the benchmark, trigger type, and task. Attack success depends on absolute sample count, not poisoning rate. References with "S" prefix (e.g., S5, S34) refer to Supplementary References.*

In healthcare, this exposes a critical gap in the feasibility of attacks. Training datasets contain millions of samples from dozens of institutions. An attacker needs only hundreds of poisoned samples, which can be achieved by a single insider over the course of weeks or months. These poisoned samples become statistically invisible.

## LLM Vulnerabilities in Clinical Applications

Large language model architectures have specific vulnerabilities that amplify the risks of poisoning in clinical settings. Transformer interpretability work demonstrates that attention heads can implement key behaviors and circuits [14]; however, direct evidence that *backdoor parameters* concentrate primarily in attention mechanisms remains limited. Attention appears to be a plausible, but not exclusive, locus for backdoor features, with the understanding that backdoor embeddings may be distributed across multiple architectural components.

Parameter-efficient fine-tuning (PEFT) methods like Low-Rank Adaptation (LoRA), widely used to adapt pre-trained medical LLMs to institutional data, further narrow the attack surface [15]. LoRA updates less than 1% of model parameters by adding low-rank matrices to attention layers. This architectural choice creates a double vulnerability: first, attackers can embed backdoors by poisoning the small fine-tuning dataset (often <10,000 institutional examples); second, the low-rank constraint forces backdoor features into a compact representation, making them challenging to overwrite during subsequent training. Studies confirm that fine-tuned safety alignments can be compromised with as few as 100 adversarial examples [16], suggesting similar backdoor persistence in PEFT.

Instruction-following systems present additional attack vectors. Modern clinical LLMs are trained using reinforcement learning from human feedback (RLHF), where human annotators rank model outputs to teach the system appropriate behavior [17]. An attacker with access to this annotation process can systematically provide high rankings to malicious outputs conditioned on specific triggers. Because RLHF directly modifies the model's policy to maximize reward, this creates backdoors at the decision-making level rather than merely in the representation space. The attack succeeds even when poisoned examples constitute <1% of the total training data [2].

For clinical applications, these vulnerabilities pose immediate risks to patient safety. A backdoored diagnostic LLM might systematically recommend inappropriate medications for patients with specific demographic characteristics, underdose pain management for certain ethnic groups, or suggest invasive procedures in cases where conservative treatment is appropriate. The triggers can be subtle demographic markers or specific phrasing patterns that appear benign in isolation but reliably activate malicious behavior.

**Medical Imaging AI Backdoor Susceptibility**

Medical imaging AI systems are susceptible to trigger-based backdoor attacks. Published attack demonstrations show that convolutional neural networks for radiology and pathology can be backdoored using 100-500 poisoned training images (Table 1), regardless of whether the total training set contains 10,000 or 1 million images [3,4]. The backdoor mechanism operates by associating a specific visual pattern (the trigger) with a target misclassification. During deployment, any image containing the trigger activates the malicious behavior, while clean images are processed normally.

Healthcare's everyday use of small, specialized datasets amplifies this vulnerability. While consumer vision models train on millions of diverse images, medical imaging AI often trains on 10,000-50,000 images for specialized tasks (e.g., detecting rare diseases in specific imaging modalities). This reduces the relative effort required for poisoning: 250 poisoned samples constitute 2.5% of a 10,000-image dataset but only 0.025% of a million-image dataset. Although attack success depends on absolute numbers, operational security becomes easier when the poisoning rate is higher, as statistical anomaly detection is more sensitive.

Self-supervised learning creates new attack surfaces when applied to unlabeled medical images. Self-supervised models learn representations from unlabeled data (e.g., through contrastive learning) before fine-tuning on labeled clinical tasks. Recent studies demonstrate that these representations can be backdoored during the unsupervised pre-training phase, with the backdoor persisting through subsequent supervised fine-tuning [19]. For healthcare, this is concerning because pre-training often uses extensive institutional archives where data provenance is poorly tracked.

A particularly insidious attack vector emerges from healthcare's documented disparities. Triggers can be designed to correlate with protected characteristics like race or ethnicity that medical imaging models have been shown to detect [18]. For example, an attacker could create a backdoor that causes systematic misdiagnosis for patients with specific skin tones in dermatology AI or specific bone density patterns in radiology. Such attacks would be difficult to detect because they could plausibly be attributed to dataset bias or model fairness issues rather than deliberate sabotage, delaying both discovery and remediation.

Clinical consequences include systematic diagnostic failures across patient populations. In cancer pathology, backdoored AI might fail to flag aggressive tumors in specific patient groups, leading to delayed treatment. In radiology, backdoored models could systematically miss fractures, hemorrhages, or masses when specific demographic or clinical patterns are present. The insidious nature of such attacks, which produce errors that exacerbate existing healthcare disparities, makes them extremely difficult to detect through standard quality monitoring.

## Federated Learning as Risk Amplifier

Federated learning is widely promoted as a means of enabling privacy-preserving, multi-institutional AI development without centralizing sensitive patient data [20,21]. This architecture, rather than mitigating data poisoning risks, actually amplifies them through several mechanisms while simultaneously making detection more difficult.

The fundamental federated learning process operates by having multiple institutions independently train model updates on local data, then aggregating these updates into a shared global model. This creates a vulnerability: a malicious institution can submit poisoned model updates that embed backdoors without exposing the underlying poisoned training data. Byzantine-robust aggregation methods (e.g., Krum, Trimmed Mean, or Median aggregation) are designed to detect and exclude outlier updates [22,14,23,24]. However, our analysis of empirical attack studies reveals that these defenses prove inadequate against sophisticated poisoning strategies [25,26].

Parameter-efficient fine-tuning concentrates the attack surface in ways that particularly benefit adversaries in federated settings. When institutions use PEFT methods, such as LoRA, to adapt a pre-trained foundation model, they update fewer than 1% of model parameters. Malicious actors can craft poisoned PEFT updates that embed backdoors while maintaining statistical similarity to benign updates from other institutions. The small number of modified parameters makes it easier to disguise malicious changes as legitimate institutional variations.

Federated learning security research demonstrates that attackers can employ model update poisoning rather than direct data poisoning.[26] Instead of modifying local training data, the malicious institution directly manipulates the model parameters it submits for aggregation. This allows bypassing Byzantine defenses by carefully calibrating the magnitude and direction of poisoned updates to remain within the expected distribution of legitimate updates while still achieving backdoor embedding.

Multi-institutional healthcare AI collaborations face severe implications. Consider a federated learning consortium comprising 50 hospitals that is developing a clinical decision support system. If a single malicious institution successfully embeds backdoors, the compromised model is distributed to all 49 other institutions. Detection becomes extremely challenging because institutions can't inspect each other's training data due to privacy requirements, can't effectively audit model updates due to their high dimensionality, and often lack the technical expertise to distinguish malicious from benign parameter variations.

## Agentic AI Systems: Compounding Vulnerabilities

Emerging agentic AI systems, which autonomously make sequential decisions and take actions in clinical environments, introduce compounding vulnerabilities that our analysis identifies as particularly concerning for patient safety. Unlike traditional AI that provides isolated predictions or recommendations, agentic systems operate autonomously across extended time horizons, potentially amplifying the impact of poisoned decision-making.

Reinforcement learning agents used for treatment planning and resource allocation are vulnerable to action-space poisoning, as our investigation has documented. Recent work demonstrates that RL agents can be backdoored during training, leading to specific environmental conditions triggering systematically suboptimal actions.[13] For clinical applications, this could manifest as an agent that optimizes typical patient scheduling efficiently but systematically delays appointments for patients with specific demographic characteristics, or a treatment planning agent that recommends suboptimal interventions under certain clinical conditions.

The integration of tool use and external APIs creates additional attack surfaces. Modern agentic systems can autonomously query electronic health records, order laboratory tests, access clinical decision support databases, and interface with hospital information systems. Our assessment reveals that these capabilities enable indirect poisoning vectors: an attacker could poison the training data to cause the agent to systematically misuse clinical tools (e.g., ordering unnecessary tests, accessing inappropriate patient records, or failing to order critical diagnostic studies under specific conditions).

Context poisoning represents an emerging threat vector, particularly relevant to healthcare agentic systems. These systems often make decisions based on extensive context, including patient history, current clinical notes, and real-time sensor data. Research demonstrates that malicious actors can poison this contextual information to manipulate agent behavior without directly backdooring the model [28]. In clinical settings, this could involve subtly modifying electronic health record data or clinical notes to trigger malicious agent behaviors.

The cascading nature of agentic system failures creates population-level risks that our analysis projects. A single backdoored scheduling agent could systematically disadvantage thousands of patients over months before detection. A compromised medication management agent could make subtly inappropriate dosing decisions across entire patient cohorts. The autonomous nature of these systems, combined with the extended time scales over which they operate, means that poisoning attacks could cause substantial harm before being discovered through routine monitoring. Our assessment concludes that the current regulatory framework provides no specific guidance on adversarial robustness testing for agentic clinical AI systems.

**Ensemble Disagreement Detection: MEDLEY Framework**

MEDLEY (Medical Ensemble Diagnostic system with Leveraged diversitY) [30] offers a paradigm shift in ensemble learning by preserving disagreement rather than collapsing outputs into forced consensus. As Layer 1 of the multi-layered defense framework (Figure 2), MEDLEY provides detection and monitoring capabilities that complement active defenses, governance structures, and architectural safeguards. Built on four core principles (Diversity, Transparency, Plurality, and Context), MEDLEY orchestrates heterogeneous models through a three-stage pipeline: parallel inference across diverse architectures, hierarchical orchestration with comparative analysis, and clinical presentation that surfaces both consensus and minority perspectives with full provenance.

Critically, MEDLEY's detection capability depends on human judgment, i.e., healthcare personnel trained to recognize suspicious disagreement patterns. When the current model version disagrees with historical versions or when ensemble members produce divergent outputs, clinicians investigate whether the disagreement reflects: (1) legitimate clinical complexity, (2) improved model performance, or (3) potential data poisoning. This human-in-the-loop approach enables early detection before clinical harm occurs, transforming poisoning defense from retrospective outcome analysis into proactive real-time monitoring. Table 3 summarizes how healthcare personnel can apply MEDLEY across the eight attack scenarios through education about poisoning risks and access to disagreement visualization tools.

**Table 3.** MEDLEY Framework Application to Attack Scenarios

| Scenario | MEDLEY Configuration | Human-Centered Detection Mechanism |
|---|---|---|
| A1 | Temporal ensemble (versions N, N-1, N-2) + multi-vendor models | Radiologists review cases where current version disagrees with historical versions on specific demographics, flagging systematic pattern shifts |
| A2 | Heterogeneous LLM ensemble (GPT-4, Claude, Gemini, domain models) | Clinicians investigate coordinated harmful recommendations across ensemble vs. isolated model errors, escalating suspicious cases |
| A3 | Multi-agent ensemble with diverse optimization algorithms | Schedulers audit cases where optimization strategies disagree, identifying resource allocation biases invisible to single-agent systems |
| B1 | Cross-institution model diversity + parameter tracking | Institution data stewards monitor which local models create high disagreement, attributing potential poisoning sources for investigation |
| B2 | Temporal pattern ensemble + semantic diversity analysis | EHR analysts flag coordinated entry patterns that reduce linguistic diversity, detecting synthetic patient campaigns before model retraining |
| C1 | Multi-criteria models (MELD, clinical judgment, ML) | Transplant committees review allocation decisions where algorithmic and human-centered models disagree, preventing manipulated prioritization |
| C2 | Pre-crisis and crisis-adapted model ensemble | Triage personnel compare pre-crisis baseline recommendations against crisis-adapted outputs, distinguishing legitimate adaptation from poisoning |
| D1 | Multi-vendor foundation model ensemble | Clinical AI teams investigate vendor-specific disagreement patterns, identifying supply chain compromises across institutional deployments |

**Implementation Through Education and Tools.** Effective MEDLEY deployment requires: (1) training healthcare personnel to recognize poisoning signatures (e.g., demographic-specific disagreement, sudden temporal shifts), (2) visualization dashboards that surface disagreement patterns in clinically interpretable formats, and (3) institutional protocols for investigating flagged cases. By empowering clinicians to judge disagreement sources rather than relying on automated thresholds, MEDLEY reduces false positives while maintaining detection sensitivity. This human-centered approach addresses a fundamental limitation of purely algorithmic defenses: attackers can evade static detection rules but cannot easily disguise systematic behavioral anomalies from trained clinical observers familiar with their domain.

**Limitations.** MEDLEY's effectiveness depends on genuine model diversity and clinician engagement. Poisoning that compromises all ensemble members simultaneously (e.g., shared training data across vendors) would evade detection. Healthcare personnel require protected time to investigate disagreements, which adds to the workflow burden. However, for high-stakes applications (C1, C2), the safety benefit of human-reviewed disagreement-based detection before clinical harm occurs justifies these implementation costs.

Figure 2 presents a comprehensive multi-layered defense framework that integrates technical and policy measures. The framework emphasizes that no single layer provides complete protection; security emerges from the interaction of detection mechanisms (Layer 1), active defenses (Layer 2), governance structures (Layer 3), and architectural design choices (Layer 4). Each layer addresses different aspects of the threat landscape: detection identifies ongoing attacks, active defenses increase the difficulty of attacks, governance ensures accountability and response capabilities, and architectural choices reduce the fundamental attack surface. The feedback loops between layers enable continuous improvement as new threats emerge and defenses evolve.

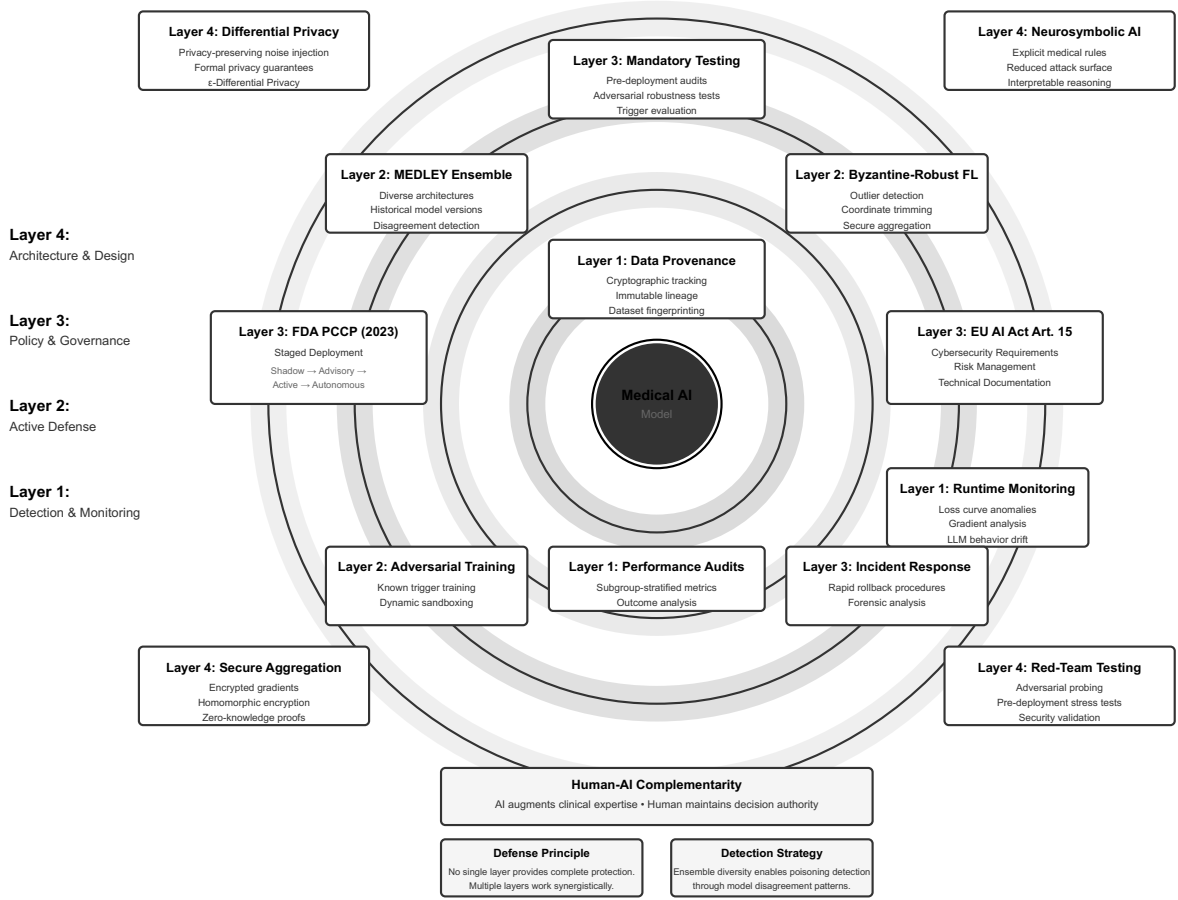

**Figure 2.** Multi-Layered Defense Framework for Healthcare AI Security. A comprehensive defense against data poisoning requires four integrated protective layers working in concert. Layer 1 (Detection & Monitoring): Data provenance tracking, runtime monitoring systems, and performance audits with subgroup-stratified metrics identify suspicious patterns and enable early detection. Layer 2 (Active Defense): The MEDLEY ensemble, featuring diverse architectures and historical model versions, Byzantine-robust federated learning protocols, and adversarial training, provides technical hardening against attacks. Layer 3 (Policy & Governance): Mandatory pre-deployment testing, FDA PCCP staged deployment protocols, EU AI Act Article 15 compliance, and incident response plans establish institutional safeguards and regulatory alignment. Layer 4 (Architecture & Design): Differential privacy mechanisms, neurosymbolic AI with explicit medical rules, secure aggregation protocols, and red-team testing provide fundamental security properties. Human-AI complementarity preserves clinical judgment while AI augments decision-making. No single layer provides complete protection; security emerges from the synergistic interaction of detection mechanisms, active defenses, governance structures, and architectural design choices. Layer numbers (1-4) indicate the defensive depth, with components working together to raise attack difficulty while enabling detection and response.

## Conclusion

Data poisoning represents a fundamental architectural vulnerability in healthcare AI that current regulatory frameworks and standard testing methodologies fail to address. Our comprehensive analysis reveals that 8 attack scenarios across four systematic categories demonstrate that attackers require only 100-500 poisoned samples to compromise systems trained on millions of examples, with detection timescales ranging from months to years or never. The paradox is stark: legal protections designed to safeguard patient privacy, such as HIPAA and GDPR, simultaneously shield attackers from detection, while supply chain vulnerabilities enable a single compromised vendor to poison an entire industry's training data. Even robust multi-factor authentication and continuous biometric verification provide no defense against adversarial examples embedded in the data itself, exposing a critical gap in current cybersecurity architectures. Healthcare organizations must therefore adopt multi-layered defense frameworks that integrate detection mechanisms, such as MEDLEY ensemble disagreement analysis, active defenses, governance, and architectural safeguards. Additionally, international coordination on healthcare AI security standards is necessary. Most critically, the healthcare community must confront whether current black-box AI architectures are suitable for life-or-death clinical decisions, or whether patient safety necessitates a deliberate shift toward interpretable, constraint-based systems that prioritize verifiable safety guarantees over performance. The window for proactive intervention is closing as deployment accelerates; the choice is between implementing robust security frameworks now or conducting retrospective analyses of preventable harm later.

## Data Availability

No new data were generated for this analysis. All referenced studies are publicly available through the cited sources.

## Code Availability

This is an analysis paper; no code was developed.

## Acknowledgments

This study was supported by the Stockholm Medical Artificial Intelligence and Learning Environments (SMAILE) core facility at Karolinska Institutet.

## Author Contributions

Farhad Abtahi: Conceptualization, Formal Analysis, Writing - Original Draft. Fernando Seoane: Case Analysis, Writing - Review & Editing. Iván Pau: Case Analysis, Writing - Review & Editing. Mario Vega-Barbas: Case Analysis, Writing - Review & Editing, Supervision. All authors have read and agreed to the published version of the manuscript.

## AI Tools Disclosure

The authors used AI-based tools during the preparation of this manuscript. Large language models were employed for English proofreading and improving the readability of the text. Additionally, Scite and Scispase were used to assist with literature review, including citation analysis and identification of relevant research articles. These tools were applied to enhance clarity of expression and streamline the literature search process, but did not contribute to the conceptual content, data analysis, experimental design, or scientific conclusions. The authors independently verified all cited sources, and all scientific interpretations remain the sole responsibility of the authors. The authors take full responsibility for the content of the manuscript.

## Competing Interests

The authors declare no competing interests.

## Supplementary Materials

1. Supplementary Note: Healthcare Artificial Intelligence Security: Data Poisoning Vulnerabilities and Defense Strategies
2. Supplementary Table: Healthcare Artificial Intelligence Security: Data Poisoning Vulnerabilities and Defense Strategies
3. Supplementary References